\documentclass[final,3p,times]{elsarticle}
\usepackage{graphicx}\usepackage{xcolor}
\newcommand{\bea}{\begin{eqnarray}}
\newcommand{\eea}{\end{eqnarray}}

\journal{Elsevier}

\begin{document}
\begin{frontmatter}

\title{Closed-form Br\"uckner $G$-matrix and nuclear matter in EFT($\not\!\pi$)}

\author[a]{Ting-Wei Pan}
\author[a,b]{Ji-Feng Yang\corref{cor1}}
\address[a]{School of Physics and Electronic Science, East China Normal University, Shanghai 200241, China}
\address[b]{State Key Laboratory of Precision Spectroscopy, East China Normal University, Shanghai 200241, China}
\cortext[cor1]{Corresponding author}
\begin{abstract}
The closed-form Br\"uckner $G$ matrix for nuclear matter is computed in the $^1S_0$ channel of EFT($\not\!\!\pi$) and renormalized in nonperturbative context. The nuclear medium environment yields additional constraints that are consistent with off-shell $T$ matrix renormalization, keeping the power counting intact and simplifying the running behaviors of the EFT couplings. With the $G$ obtained we computed the energy per particle for neutron and symmetric nuclear matter to demonstrate the physical relevance of certain 'physical' parameters that arise from nonperturbative renormalization. We also explored the pairing phenomenon in $^1S_0$ channel by examining the poles of the closed-form $G$ matrix with a given density, where again the physical relevance of the same set of physical parameters are clearly illustrated.
\end{abstract}
\begin{keyword}
Nuclear matter\sep Pionless EFT\sep Nonperturbative renormalization
\end{keyword}
\end{frontmatter}

\section{Introduction}EFT has become a main tool for quantitative study of many issues in nuclear physics, both for few body physics and for many body issues. Among the various EFTs for nuclear physics, the pionless EFT or EFT$(\not\!\!\pi)$ is particularly useful in illustrating many important issues\cite{review1,review2,review3} for its simple structure that facilitates both perturbative and nonperturbative studies.

EFT$(\not\!\!\pi)$ has been employed in our previous works\cite{YH,AOP,1806} for demonstrating alternative solution of $T$-matrices in nonperturbative contexts, namely closed-form solutions and the corresponding nonperturbative scenario of nuclear dynamics in very low energy situation. In contrast to the known 'perturbative' schemes like KSW\cite{KSW}, it is shown that a novel scenario of renormalization of the simpler theory is tractable in the presence of tight constraints imposed by the closed form of the $T$-matrices, with a conventional realization of EFT power counting that is naturally compatible with unnatural scattering behaviors between nucleons\cite{YH,AOP,1806,epl85,epl94}.

In this report, we wish to extend our approach to many-body contexts and to show that the alternative scenario of renormalization in nonperturbative contexts also works in such many body issues. To this end, we consider the closed-form Br\"uckner $G$ matrices in uncoupled channels of EFT$(\not\!\!\pi)$. The first computation of Br\"uckner $G$ matrix in EFT$(\not\!\!\pi)$ has been done twenty years ago in Ref.\cite{krippa}. Actually, the pionless EFT has been perturbatively applied to many-body fermi gas in a number of papers almost two decades ago, see, e.g., Refs\cite{HF,Steele}. We wish to extend the EFT framework to nonperturbative context with adapted renormalization. From our presentation, it is obvious that Br\"uckner $G$ matrices could be readily obtained from the closed-form $T$ matrices by merely replacing the density-independent integrals by the density-dependent ones. Thus, the extension seems to be an easy job. The presence of medium does bring us additional constraints in the closed-form $G$ matrix, which imply additional 'physical' or renormalization group invariant parameters to be independently fixed, reducing the number of running parameters. Naturally, there should be more physical inputs to fix these parameters in the presence of medium. As long as we work with EFT at a given order, the number of parameters to be fixed remain finite, even in a nonperturbative context. In this sense, the EFT spirit is feasible even beyond the most familiar perturbative formulations, as long as the traditional perturbative wisdom is properly updated.

This report is organized as below: In Sec. 2 the necessary theoretical set up is presented and the closed-form Br\"uckner $G$ matrix in $^1S_0$ channel is computed via Bethe-Goldstone equation with the potential given at the first two truncation orders. The renormalization of the closed-form of $G$ matrix is described in Sec. 3 through exploiting the tight constraints imposed by the closed-form of $G$ with the running couplings given as byproducts. The power counting for such EFT description is also described in Sec. 3. Sec. 4 is devoted to the calculation of energy per nucleon in $^1S_0$ channel for demonstration of the physical relevance of the parameters derived from the tight constraints. Sec. 5 is devoted to a demonstration of the pairing phenomenon with the closed-form $G$ matrix. The discussion and summary is given in Sec 6.
\section{Closed-form Br\"uckner $G$ matrices}The pionless EFT for low energy $NN$ interactions is characterized by the following local lagrangian:\bea\mathcal{L}_{(\not\pi)}=\bar{N} \left(i\partial_0+\frac{\nabla^2}{2M_N}\right)N-\frac{1}{2}C_s\left(\bar{N}N\right)^2-\frac{1}{2}C_t\left(\bar{N}\vec{\sigma}N\right)^2+\cdots,\eea from which we could naturally read off the two-body (energy-independent) pionless potential of $^1S_0$ channel truncated at the first two orders:\bea\mathcal{O}(Q^0):\ V(q,q^\prime)=C_{s;0};\quad\quad\mathcal{O}(Q^2):\ V(q,q^\prime)=C_{s;0}+C_{s;2}\left(q^2+{q^\prime}^2\right).\eea

It is known that the in-medium interactions for nuclear system could be described by the Br\"uckner $G$ matrix which in turn satisfies the following Bethe-Goldstone (BG) equation in uncoupled channels\cite{BG,Bishop}\bea G(q,q^\prime;p_F)=V(q,q^\prime)+M_N\int\frac{k^2dk}{2\pi^2}V(q,k)\frac{\theta(k-p_F)}{M_NE-k^2+i0^+}G(k,q^\prime;p_F).\eea Here $E$ denotes the total energy of the scattering pair of nucleons in center of mass frame, $p_F$ the Fermi momentum. Obviously, this BGE shares the same structure with the Lippmann-Schwinger equation for $NN$ scattering $T$ matrix except $\theta(k-p_F)$, and directly reduces to the latter in the zero-density limit. Moreover, just like the case of $T$ matrix, a closed-form $G$ matrix is also feasible in pionless EFT because the potential is separable.

Let us illustrate with the $^1S_0$ channel for convenience. Extension to higher channels is a straightforward matter. In analogy to our previous studies using pionless EFT\cite{YH,AOP}, let us introduce the column vector\bea U(q)=\left(\begin{array}{c}1\\q^2\end{array}\right)\eea at truncation order $\mathcal{O}(Q^2)$, to recast $V$ and $G$ into the following factorized form\bea V(q,q^\prime)&=& U^T(q)\lambda U(q^\prime),\quad G(q,q^\prime;p_F)=U^T(q)\gamma U(q^\prime),\\\lambda&\equiv&\left(\begin{array}{cc}C_{s;0}&C_{s;2}\\C_{s;2}&0\\ \end{array}\right),\ \gamma\equiv\left(\begin{array}{cc}\gamma_{11}&\gamma_{12}\\\gamma_{21}&\gamma_{22}\\\end{array}\right).\eea Then the BG equation reduces to the following algebraic one whose solution is direct to find\bea\gamma&=&\lambda+\lambda\cdot\tilde{\mathcal{I}}\cdot\gamma\rightarrow\gamma=\left(1-\lambda\cdot\tilde{\mathcal{I}}\right)^{-1}\cdot\lambda, \eea with\bea\tilde{\mathcal{I}}&\equiv&\left(\begin{array}{cc}-\tilde{\mathcal{I}}_0&\tilde{\mathcal{I}}_2\\\tilde{\mathcal{I}}_2&\tilde{\mathcal{I}}_4\\\end{array}\right).\eea It is easy to see that the renormalization of $G$ hinges upon the subtraction of the divergences residing in the matrix $\tilde{\mathcal{I}}$, which is defined and parametrized in the \ref{appa}.

Next, we could easily obtain the closed-form $G$ matrix by sandwiching $\gamma$ between the vectors $U^T$ and $U$. For studying the nuclear matter, we need the on-shell ($p=\sqrt{M_NE}$) $G$ matrix which reads at order $\mathcal{O}(Q^2)$\bea&&\frac{1}{G(p;p_F)}=\frac{\tilde{N}_{s;0}}{\tilde{D}_{s;0}+\tilde{D}_{s;1}p^2}+\tilde{\mathcal{I}}_0,\\& &\tilde{N}_{s;0}\equiv\left(1-C_{s;2}\tilde{J}_3\right)^2,\ \tilde{D}_{s;0}\equiv C_{s;0}+C^2_{s;2}\tilde{J}_5,\ \tilde{D}_{s;1}\equiv C_{s;2}\left(2-C_{s;2}\tilde{J}_3\right),\eea and in analogy to the zero-density case, it is easy to verify the following linear constraints\bea\label{ND1}\tilde{N}_{s;0} +\tilde{D}_{s;1}\tilde{J}_3=1,\ \tilde{D}_{s;0}\tilde{J}_3+\tilde{D}_{s;1}\tilde{J}_5=C_{s;0}\tilde{J}_3 +2C_{s;2}\tilde{J}_5.\eea

The on-shell $G$ matrix at leading order $\mathcal{O}(Q^0)$ is even simpler and coincides with the off-shell case:\bea\frac{1}{G(p;p_F)}=\frac{1}{C_{s;0}}+\tilde{\mathcal{I}}_0=\frac{1}{G(q,q^\prime;p_F)},\eea making it distinctive to all the higher orders in $^1S_0$ channel.

In next section, the tight constraints of the closed-form $G$ matrix will be exploited to obtain the renormalized (running) couplings, following exactly the same procedure that has been developed in Refs.\cite{YH,AOP,1806}.
\section{Tight constraints and renormalization}
\subsection{Running couplings from closed-form $G$ in $^1S_0$ channel}To proceed, we introduce the following renormalization group invariant ratios that are now density dependent:\bea\tilde{\alpha}_{s;0}\equiv\tilde{N}_{s;0}^{-1}\tilde{D}_{s;0},\ \tilde{\alpha}_{s;1}\equiv\tilde{N}_{s;0}^{-1}\tilde{D}_{s;1},\eea then the on-shell $G$ matrix could be recast into the following renormalization group invariant form\bea\frac{1}{G(p;p_F)}=\frac{1}{\tilde{\alpha}_{s;0}+\tilde{\alpha}_{s;1}p^2}+\tilde{\mathcal{I}}_0,\ \frac{1}{T(p)}=\frac{1}{G(p;p_F=0)}=\frac{1}{\alpha_{s;0}+\alpha_{s;1}p^2}+\mathcal{I}_0,\eea where\bea\label{alphazd}\alpha_{s;0}\equiv\tilde{\alpha}_{s;0}(p_F=0) =\frac{C_{s;0}+C_{s;2}^2J_5}{(1-C_{s;2}J_3)^2},\ \alpha_{s;1}\equiv\tilde{\alpha}_{s;1}(p_F=0)=\frac{2C_{s;2}-C_{s;2}^2J_3}{(1-C_{s;2}J_3)^2},\ \mathcal{I}_0\equiv J_0+i\frac{M_N}{4\pi}p.\eea

As in the zero-density case\cite{YH,AOP,1806}, it is easy to see that $\tilde{\mathcal{I}}_0$ becomes renormalization group invariant in $G$ beyond leading order of truncation, so is $J_0=\Re\{\tilde{\mathcal{I}}_0(p_F=0)\}$. Moreover, the presence of density further constrains $J_3$ to be also renormalization group invariant, see \ref{appb}. This is not totally unanticipated if we further explore the renormalization of the off-shell $T$ matrix. For example in the $^1S_0$ channel at order $\mathcal{O}(Q^2)$, to renormalize the off-shell $T$ matrix in closed-form, the off-shell piece $\check{\delta}_{S}=-\mathcal{I}_0C^2_{s;2}$ (c.f. \cite{1806}) must conspire with $N_{s;0}$ to constitute  a further renormalization group invariant $\check{\delta}_{S}/N_{s;0}$, which equals to $\tilde{\mathcal{I}}_0(p_F=0)\sigma^2_{s;2}$ with $\sigma_{s;2}$ defined in \ref{appb}.

Working out all the renormalization group invariants, one could proceed to solve for the running couplings in the following two routes: One is to invert the expressions of appropriate renormalization group invariants to obtain running couplings; Another is to invert $\tilde{\alpha}_{s;0}$ and $\tilde{\alpha}_{s;1}$ for couplings, then to plug in more renormalization group invariance to cancel out the apparent density dependence. In either route, we need to first work out the highest order couplings, then go to lower orders, in perfect accordance with the general notion established in our previous studies of the closed-form $T$ matrices in pionless EFT that the renormalization of lower order couplings could be affected by higher order couplings, not the reverse\cite{YH,AOP,1806}. In our view, this should be a very 'natural' feature for any EFT formulation of subatomic and atomic systems. Unfortunately, it has not been much manifested or appreciated in literature.

Now, from Eqs.(\ref{sig}) and (\ref{identity}) in \ref{appb}, we have\bea\label{RC2}{C}_{s;2}=\frac{\sigma_{s;2}}{1+\sigma_{s;2}{J}_3}=\frac{\sigma_{s;2}^2}{\alpha_{s;1} -\sigma_{s;2}}.     \eea Interestingly, $C_{s;2}$ is now completely fixed or renormalization group invariant, a byproduct of the additional constraints due to medium background (or, renormalization of off-shell $T$ matrix). Next, we can simply find from Eq.(\ref{alphazd}) that\bea\label{RC0}C_{s;0}=\frac{\alpha_{s;0}-\sigma_{s;2}^2J_5(\mu)}{(\alpha_{s;1} -\sigma_{s;2})^2}\sigma_{s;2}^2, \eea which could also be found by using the second relation in Eq.(\ref{ND1}).

As to the second route, we simply make use of the published results of the zero-density case\cite{YH,1806} and the replacement $[J_{\cdots}]\rightarrow[\tilde{J}_{\cdots}]$: \bea\label{RCtilde}\tilde{C}_{s;2}=\tilde{J}^{-1}_3\left[1-\left(1+\tilde{\alpha}_{s;1}\tilde{J}_3\right)^{-1/2}\right],\ \tilde{C}_{s;0}=\frac{\tilde{\alpha}_{s;0}}{1+\tilde{\alpha}_{s;1}\tilde{J}_3}-\tilde{J}_5\tilde{J}^{-2}_3\left[1-\left(1+\tilde{\alpha}_{s;1}\tilde{J}_3\right)^{-1/2}\right]^2.\eea Then, with the help of Eqs.(\ref{al0}) and (\ref{identity}), all the seemingly $p_F$-dependence cancel out and reproduce the results in Eqs.(\ref{RC2}) and (\ref{RC0}), see \ref{appc}. Going down to the leading order, the solution of $C_{s;0}$ would jump to the well-known KSW running as $J_0$ is mixed with $C_{s;0}$ and hence runs, ${C}_{s;0}=\frac{\alpha_{s;0}}{1+\alpha_{s;0}J_0(\mu)}$, a very accidental situation that is in sheer contrast to all the higher orders' situation\cite{1806}.

The foregoing study of the $^1S_0$ channel at $\mathcal{O}(Q^2)$ can be the guiding example for higher channels and higher orders. As an example, the uncoupled $P$ channels will be explored in \ref{appd} and \ref{appe}.
\subsection{'Effective' density-dependent couplings}Before leaving this section, we note in this subsection that the density-dependent ratios $[\tilde{\alpha}_{L;i}]$ actually serve as 'effective' couplings in the on-shell $G$ matrices in $L$ channel, whose dependence upon density is theoretically physical. Bringing in the values of the ratios $[{\alpha}_{L;i}]$ that could be reexpressed in terms of ERE parameters or other physical inputs, one could acquire the physical density-dependence of such (on-shell) 'effective' couplings. Let us illustrate such dependence with the $^1S_0$ channel at order $\mathcal{O}(Q^2)$, for the purpose of comparison with Ref.\cite{krippa}.

To this end, we define the 'effective' density-dependent couplings of $^1S_0$ channel as below (C.f.\ref{appb}):\bea\tilde{C}^{(\texttt{\tiny eff})}_{s;0}(p_F)\equiv\tilde{\alpha}_{s;0}=\frac{\alpha_{s;0}+\sigma^2_{s;2}\kappa_5}{\left(1-\sigma_{s;2}\kappa_3\right)^2},\quad\tilde{C}^{(\texttt{\tiny eff})}_{s;2}(p_F)\equiv\frac{1}{2}\tilde{\alpha}_{s;1}=\frac{\alpha_{s;0}-\sigma^2_{s;2}\kappa_3}{2\left(1-\sigma_{s;2}\kappa_3\right)^2},\eea with which we could write\bea\frac{1}{G(p;p_F)}=\frac{1}{\tilde{C}^{(\texttt{\tiny eff})}_{s;0}(p_F)+2\tilde{C}^{(\texttt{\tiny eff})}_{s;2}(p_F)p^2}+\tilde{\mathcal{I}}_0.\eea Here, we note in passing that as the dimensionless factor $\sigma_{s;2}\kappa_3\sim0.21{p^3_F}/\Lambda^3_{\not\pi}$ is small after taking into account the fact that $p_F$ is no larger than $\Lambda_{\not\pi}$ (upper scale of EFT($\not\!\pi$)) for our EFT approach to make sense, the formal singularity in the 'effective' couplings due to $\left(1-\sigma_{s;2}\kappa_3\right)^{-2}$ does not materialize within the realm of EFT($\not\!\pi)$, making their $p_F$-dependence actually a mild one. Therefore, the dominating influence of density background comes from the 'physical' parameter $\tilde{\mathcal{I}}_0=J_0+\frac{M_N}{4\pi^2}\left(p\ln\frac{p_F+p}{p_F-p}-2p_F\right)$ for $p\leq p_F$. Actually, when $p_F$ is around $\Lambda_{\not\pi}$, $G(p;p_F)$ is still regular and exclusively controlled by $\tilde{\mathcal{I}}_0$: $1/G(p;p_F=\Lambda_{\not\pi})=\tilde{\mathcal{I}}_0$, another merit in using the closed form of $G$ matrix.
\subsection{Power Counting}In this subsection, we describe the power counting involved in the $G$ matrix, namely the scaling laws for the couplings, parameters from integrals and the renormalization group invariant ratios.

In our works\cite{YH,AOP,1806,epl85,epl94}, a natural scenario of EFT have been explored and tested in order to be able to describe unnaturally large $^1S_0$ scattering length with closed-form $T$ matrix. We have demonstrated that with a large renormalization group invariant $J_0$, the following simple rules of power counting could realize the large $^1S_0$ scattering length that incorporates naturally sized couplings:\bea&&{C}_{L;2i}\sim\frac{4\pi} {M_N\Lambda^{2i+1}_{\not\pi}},\ i=0,1,2,\cdots;\ {J}_0\sim\frac{M_N\Lambda_{\not\pi}}{4\pi}\sim{C}^{-1}_{s;0},\ {J}_{2l+1}\sim\frac{M_NQ^{2l+1}}{4\pi},\ l=1,2,\cdots,\eea with $Q\sim\mu\sim\epsilon\Lambda_{\not\pi}$. For $p_F$, we note that Fermi momentum is actually an external scale for EFT that only enter the game via loop integrals to represent the many-body environment provided by medium. Thus due to their entrance through the EFT integrals, we can count it as either $p_F\sim\Lambda_{\not\pi}$ or $p_F\sim\epsilon\Lambda_{\not\pi}$, depending on practical issues.

Then we have the following scaling laws for the renormalization group invariants\bea&&\alpha_{L;i}\sim\tilde{\alpha}_{L;i}\sim\eta\frac{4\pi}{M_N\Lambda^{2i+1}_{\not\pi}}\sim\eta{C}_{L;2i},\ \beta_{L;i}\sim\tilde{\beta}_{L;i}\sim\frac{1}{\Lambda^{2i}_{\not\pi}},\ \sigma_{L;2i}\sim {C}_{L;2i},\ i=0,1,2,\cdots\eea where $\eta=1$ for diagonal couplings and $\eta=2$ for off-diagonal couplings. Usually, $\epsilon\sim1/4$ for realistic situation, see refs.\cite{AOP,1806,epl85,epl94} for details.
\section{Energy per particle}It is known that an uncoupled partial wave (say $^1S_0$) $G$ matrix element contributes to ground state energy per particle for a system of $A$ nucleons with density $\rho=gp^3_F/(6\pi^2)$ as below\cite{krippa}\bea\frac{E_{g.s.}}{A}=\frac{3}{5}\frac{p^2_F}{2M_N}+\frac{g(g-1)}{2\rho}\int_F\frac{8d^3P} {(2\pi)^3}\frac{d^3p}{(2\pi)^3}G(p;p_F),\eea with $g=(2s+1)(2T+1)$ being the spin-isospin degeneracy and $P$ the average momentum of the incoming particles. In this report, the on-shell $G$ matrix will be partially fixed with ERE factors, the residual parameters like $J_0$ and $\sigma_{s;2}$ will be assigned with several choices to demonstrate their physical relevance. We also note in passing that the on-shell $G$ matrix becomes real once $p$ is below Fermi surface: $p\leq p_F.$

To fix the $G$ matrix in terms of ERE parameters as far as possible, we first putting that $p_F=0$ to go back to the $T$ matrix that is directly related to ERE as below\bea\Re\left\{ \frac{1}{T}\right\}=\Re\left\{\frac{1}{G(p_F=0)}\right\}=-\frac{M}{4\pi}p\cot\delta_s(p)=-\frac{M}{4\pi}\left\{-\frac{1}{a_{s}}+\frac{1}{2}r_{e;s}p^2+\sum_{k=2}^\infty{v}_{k} p^{2k}\right\}.\eea Then we have at leading order a very simple form of $G$\bea&&\frac{1}{a_s}=\frac{4\pi}{M_N}\left(\frac{1}{C_{s;0}}+J_0\right):\quad\frac{1}{G(p;p_F)}=\frac{M_N}{4\pi{a}_s}-\kappa_1+\frac{M_Np}{4\pi^2}\ln\frac{p_F+p}{|p_F-p|}.\eea Obviously, the on-shell $G$ matrix in $^1S_0$ channel is totally determined by the scattering length apart from $p$ and $p_F$ at leading order where $J_0$ is a running scale, an accidental situation that actually furnaces the KSW scheme\cite{1806}. However, it is no loner true once one goes beyond the leading order of EFT truncation. At order $\mathcal{O}(Q^2)$, we have a rather involved closed-form $G$\bea&&\frac{1}{a_s}=\frac{4\pi}{M_N}\left(\frac{1}{\alpha_{s;0}} +J_0\right),\ r_{e;s}=\frac{8\pi}{M_N}\frac{\alpha_{s;1}}{\alpha^2_{s;0}}:\nonumber\\&&\label{G2}\frac{1}{G(p;p_F)}=\frac{\left(1-\sigma_{s;2}\kappa_3\right)^2}{\left(\frac{M_N}{4\pi a_s}-J_0\right)^{-1}+\sigma_{s;2}^2\kappa_5+\left[\frac{M_Nr_{e;s}}{8\pi}\left(\frac{M_N}{4\pi a_s}-J_0\right)^{-2}-\sigma_{s;2}^2\kappa_3\right]p^2}+J_0-\kappa_1+\frac{M_Np}{4\pi^2} \ln\frac{p_F+p}{|p_F-p|}.\eea Here, the nontrivial dependence of $G$ matrix upon $J_0$ and $\sigma_2$ besides ERE factors are evident.

With the $G$ matrix given above, we may then proceed in the following two routes: A). Taylor expand $G$ in terms of powers of $p^2$ and then perform the integration analytically, hence forth called 'trimmed $G$ matrix'; B). Perform the integration numerically with the closed-form $G$ matrix.

A). First at leading order we have\bea{G(p;p_F)}=\frac{4\pi^2{a}_s}{M_N(\pi-2a_sp_F)}+\mathcal{O}(p^2),\eea which again is totally fixed by scattering length. At order $\mathcal{O}(Q^2)$, the $G$ matrix could not be totally fixed with ERE factors of the corresponding order due to the presence of $J_0$ and $\sigma_{s;2}$. However, if we count $p_F$ like $p$ also as $Q$, then the terms like $\sigma_{s;2}\kappa_3$ and $\sigma^2_{s;2}$ will be negligible and the nontrivial dependence upon $J_0$ will also be suppressed up to $\mathcal{O}(p^4,p^3_F)$\bea\label{G1S0-Q2}G(p;p_F)=\frac{4\pi^2{a}_s}{M_N(\pi-2a_sp_F)}\left\{1+\frac{\pi{a}_s}{\pi-2a_sp_F}\left(\frac{r_{e;s}}{2}-\frac{2}{\pi{p}_F}\right)p^2\right\} +\mathcal{O}(p^4,p^3_F).\eea The closed-form of on-shell $G$, as listed above in Eq.(\ref{G2}), is definitely dependent upon the values of $J_0$ and $\sigma_{s;2}$, whose physical relevance will be numerically demonstrated below.
\begin{figure}[t]\begin{center}\resizebox{10cm}{!}{\includegraphics{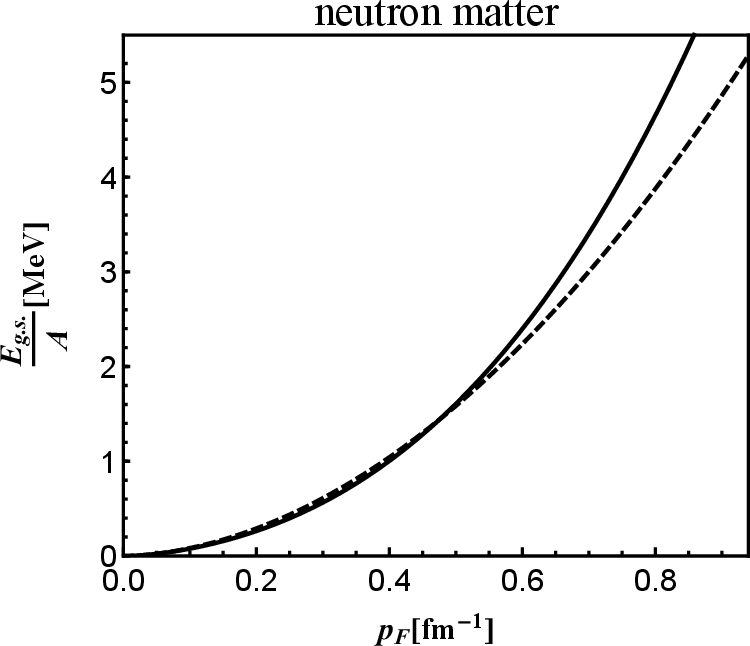}\quad\quad\quad\quad\includegraphics{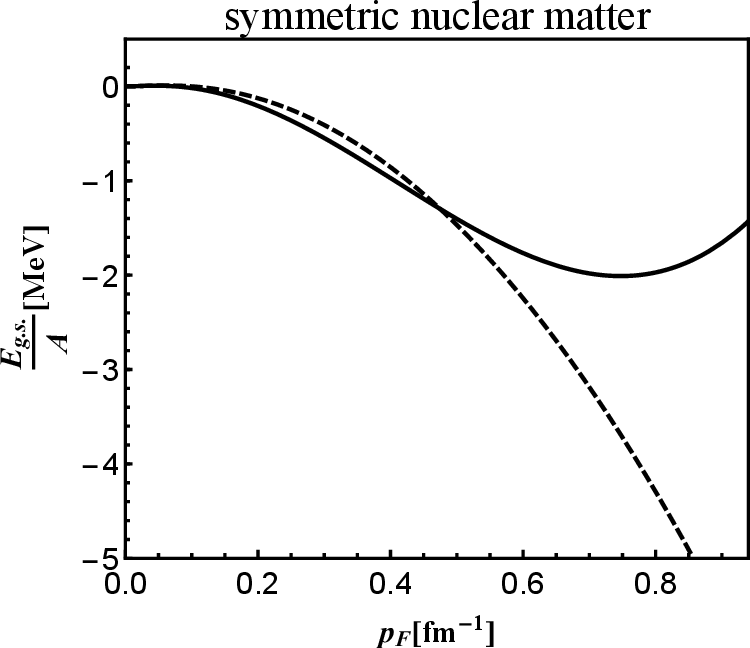}}\caption{Energy per particle in $^1S_0$ from the analytically 'trimmed $G$ matrix' that retains scattering length $a_s$ only (dashed line), and that retains scattering length $a_s$ and effective range $r_{e;s}$ (solid line).}\end{center}\end{figure} Now the energy per particle of nuclear matter at leading order of EFT truncation reads\bea\label{EpA_py_a}\frac{E_{g.s.}}{A}=\frac{p^2_F}{M_N}\left[\frac{3}{10}+\frac{(g-1)a_sp_F}{3(\pi-2a_sp_F)}+\mathcal{O}(p^3_F)\right],\eea in agreement with ref.\cite{Steele}. Beyond the leading order of EFT truncation, we have\bea\label{EpA_py_are}\frac{E_{g.s.}}{A}=\frac{p^2_F}{2M_N}\left[\frac{3}{5}+\frac{2(g-1)a_sp_F} {3(\pi-2a_sp_F)}+\frac{(g-1)\pi a^2_sp^2_F}{5(\pi-2a_sp_F)^2}\left(\frac{r_{s;e}p_F}{2}-\frac{2}{\pi}\right)+\mathcal{O}(p^5_F)\right].\eea The results given in  Eq.(\ref{EpA_py_a}) and Eq.(\ref{EpA_py_are}) are plotted in Fig.1 in a wider range of $p_F$.

Evidently, the denominator $1/(\pi-2a_sp_F)$ makes the formulae given by Eq.(\ref{EpA_py_a}) and Eq.(\ref{EpA_py_are}) better behaved than the perturbative one that is only valid when $p_F<\frac{\pi}{2a_s}\approx13.1$ MeV, a rather narrow range because of the large scattering length $a_s$, which is in support of employing nonperturbative framework to deal with nuclear matter within pionless EFT. The saturation behavior for symmetric nuclear matter ($g=4$) tends to show up only after the effective range is included (c.f. Eq.(\ref{EpA_py_are})), in other words, the leading order $NN$ interaction, could not harbor the saturation phenomenon.

B). To proceed in the second route, we will parametrize $\sigma_{s;2}$ in terms of $\alpha_{s;1}$ and $x$ as $\sigma_{s;2}=\frac{\alpha_{s;1}}{2+x}$ with $x\sim\epsilon$ due to the relation of (\ref{RC2}) and the power counting given in Sec. 4. We will demonstrate with the choices that $\Lambda_{J_0}=\frac{4\pi}{M_N}J_0$: 138MeV ($\sim\Lambda_{\not\pi}$) and 35 MeV ($\sim\epsilon\Lambda_{\not\pi}$) and $x=\pm0.2$. To show our main points, we simply work with the ERE data from proton-neutron $^1S_0$ scattering, which could be seen as a 'realistic' situation of exact isospin symmetry, i.e., a world where the electromagnetic interaction is completely turned off.

It is obvious from the curves in Fig.2 and Fig.3 that the numerical results obtained with closed-form $G$ conform to the approximate analytical results described by Eq.(\ref{EpA_py_are}) in the zero-density end, and that the value of $J_0$ matters a lot in the closed-form $G$ as a smaller value of $J_0$ will yield more pathological energy curves against $p_F$, i.e., a pole like behaviour around $p_F=0.659$fm$^{-1}$ for the $\Lambda_{J_0}=35$MeV and $x=0.2$ case (around $p_F=0.629$fm$^{-1}$ for $\Lambda_{J_0}=35$MeV and $x=-0.2$ case), namely, it is a physical parameter rather than a running parameter. The reason goes as below: In pionless EFT, the typical scale of energy per particle should be no greater than $\varepsilon\equiv\Lambda_{\not\pi}^2/M_N\approx20$MeV, thus the foregoing pole-like behavior is unacceptable. However, such violation merely comes from the choice $\Lambda_{J_0}=35$MeV, while the choice $\Lambda_{J_0}=138$MeV is free of such violation even with $p_F\sim0.9$fm$^{-1}$ as is evident in Fig.2 and Fig.3. Thus, the pole-like behavior demonstrated in Fig. 2 and 3 simply expresses that $J_0$ is a physical rather than an ordinary running parameter in the closed-form $G$ matrix of pionless EFT. Moreover, different choices of $J_0$ yield more consequential results, for example, a smaller $J_0$ would magnificently narrow down the distribution of poles in $G$ against density and hence reduce the 'space' of the formation of 'Cooper pair' of nucleons in nuclear matter, which will be discussed in next section.
\begin{figure}[t]\begin{center}\resizebox{16cm}{!}{\includegraphics{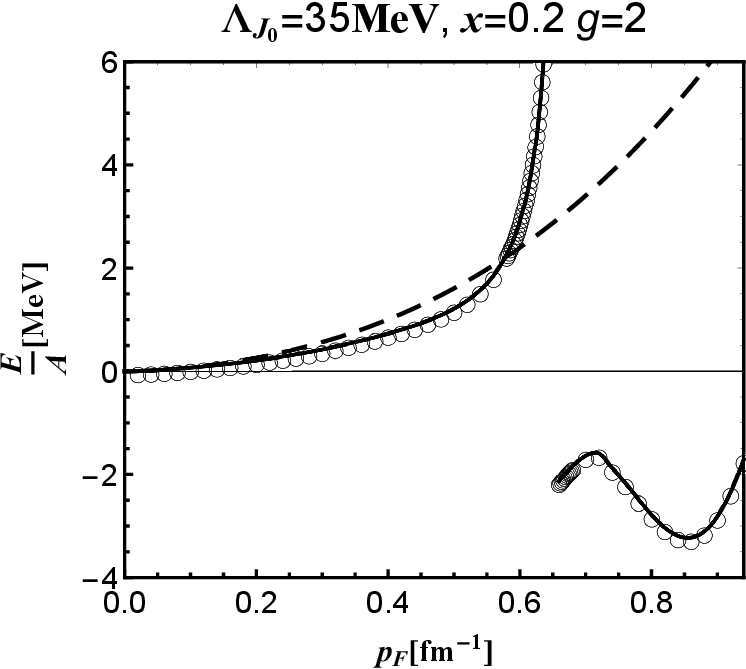}\includegraphics{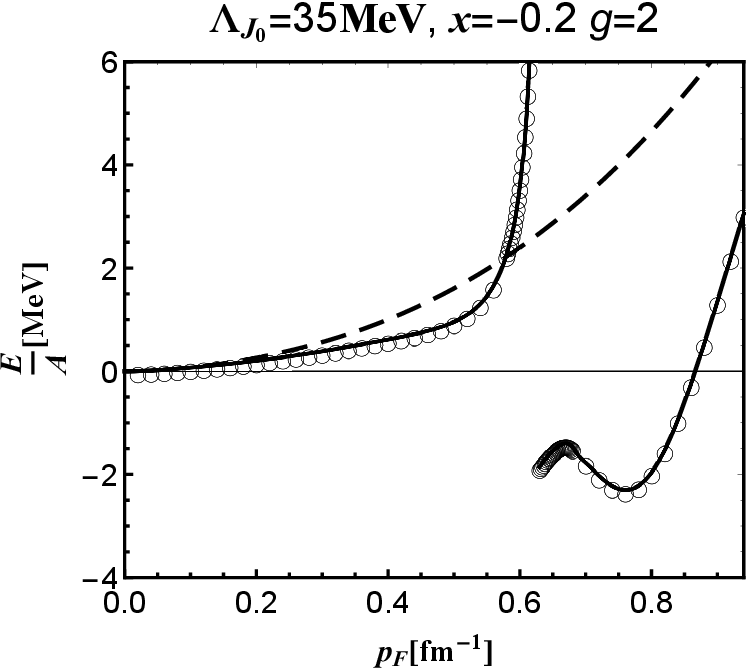}\quad\includegraphics{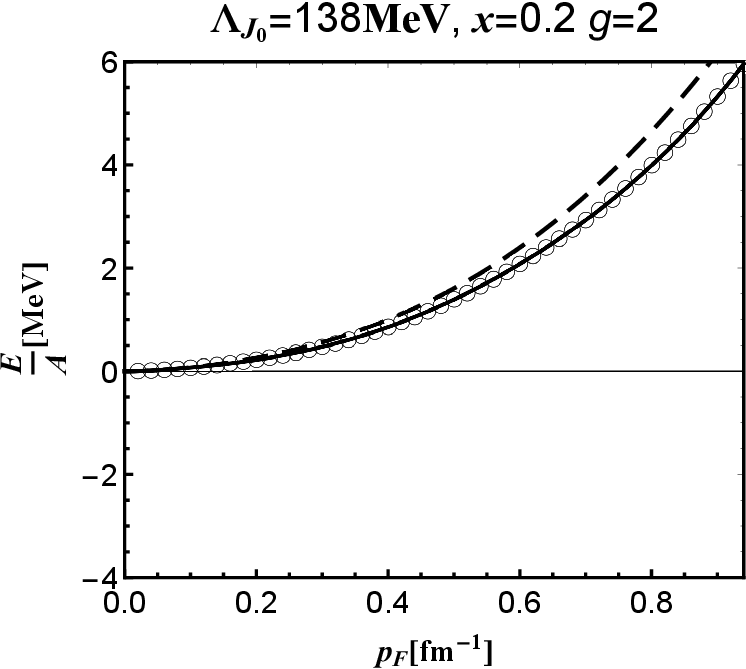} \includegraphics{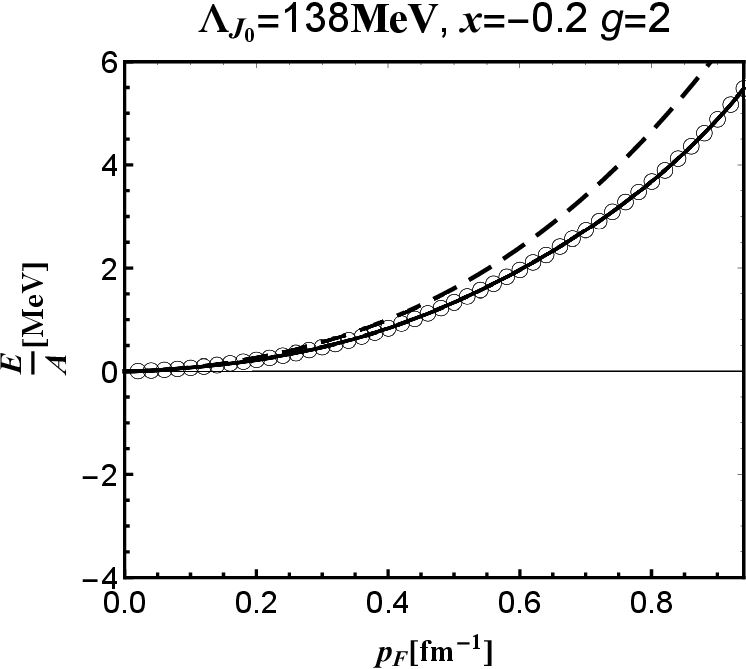}}\caption{Energy per particle of neutron matter ($^1S_0$): dashed lines are the analytical results (solid line in the left panel of Fig.1), i.e., calculated with the 'trimmed $G$ matrix' retaining $a_s$ and $r_{e;s}$, solid lines with open circles are the numerical results from direct integration with $G$ matrix without any further analytical treatment.}\end{center}\end{figure}
\begin{figure}[t]\begin{center}\resizebox{16cm}{!}{\includegraphics{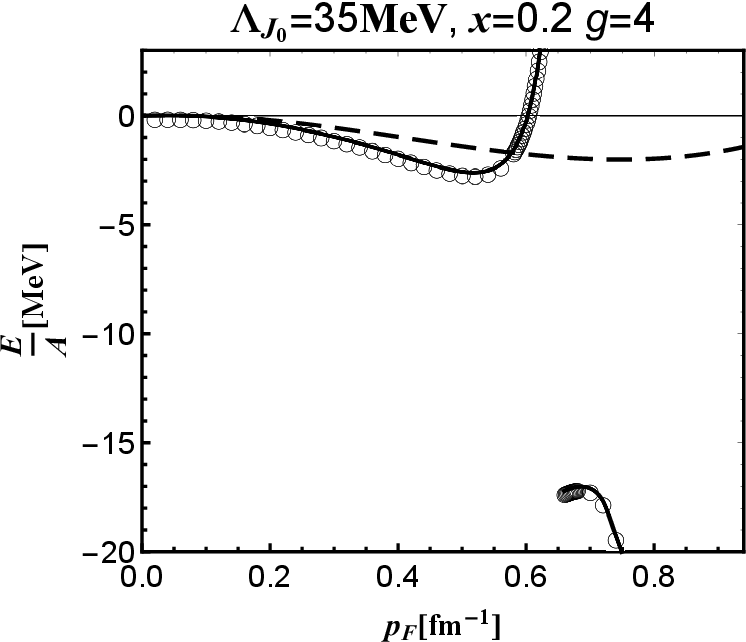}\includegraphics{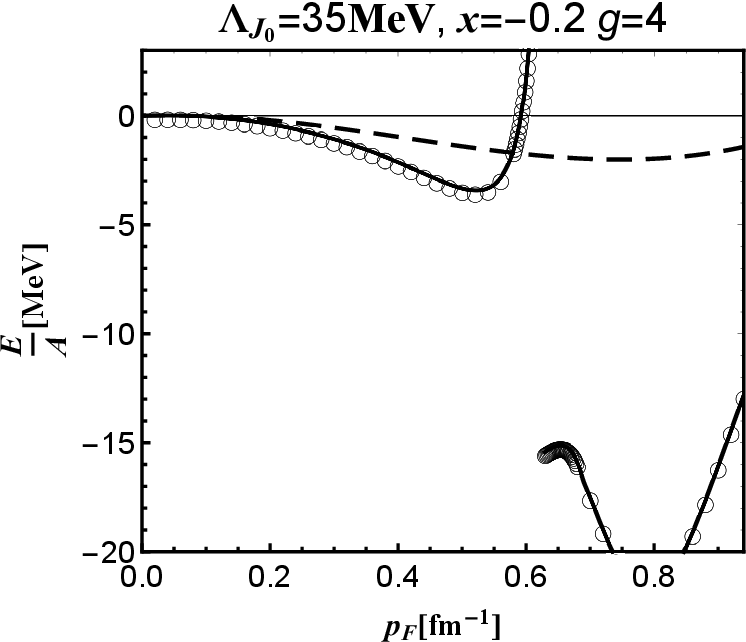}\quad\includegraphics{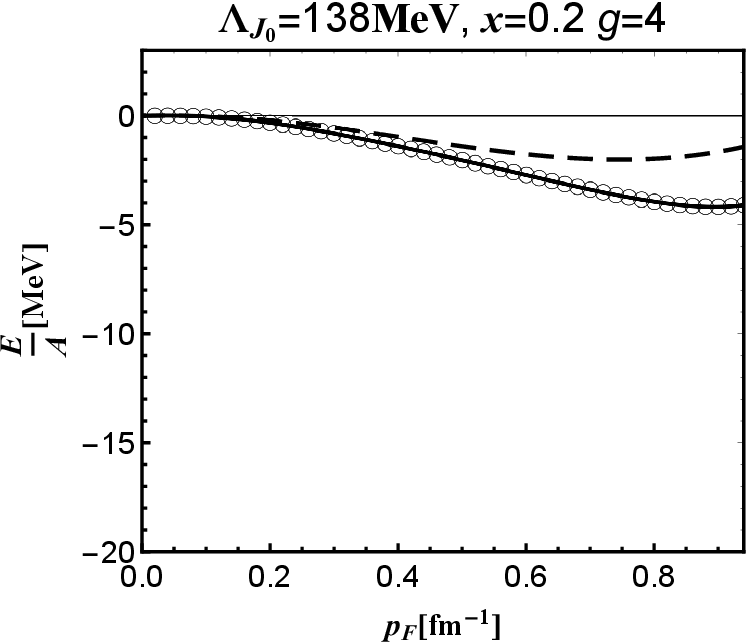} \includegraphics{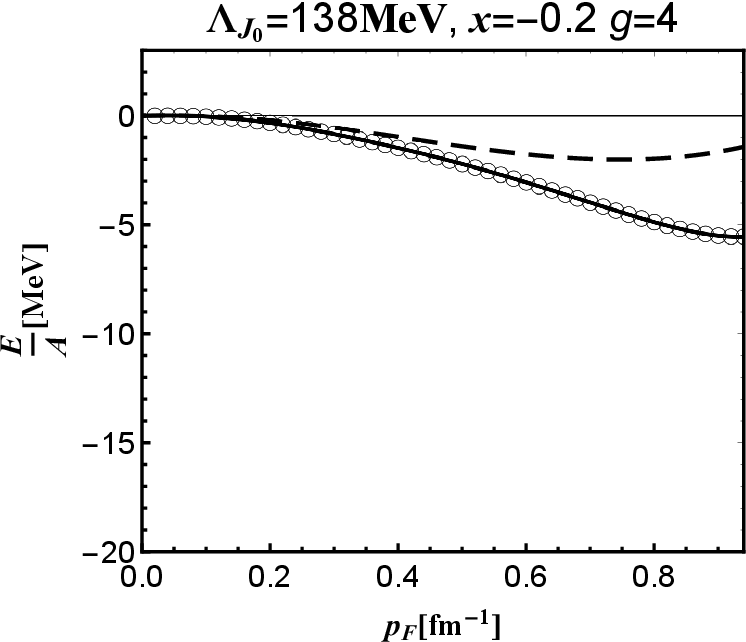}}\caption{Energy per particle of nuclear matter ($^1S_0$): dashed lines are the analytical results (solid line in the right panel of Fig.1), i.e., calculated with the 'trimmed $G$ matrix' retaining $a_s$ and $r_{e;s}$, solid lines with open circles are the numerical results from direct integration with $G$ matrix without any further analytical treatment.}\end{center}\end{figure}
\section{Pairing and poles in $G$ matrix}It is well known that nucleons tend to pair up as 'Cooper pairs' in nuclei or nuclear matter. As $G$ matrix is real as long as the on-shell momentum is below Fermi surface, here we try to establish the pairing of nucleons by examining the existence of real poles of $G$ matrix (in an attractive channel) in terms of the on-shell momentum of the incoming nucleon\cite{Schmidt,Alm,Sedra} just below a given Fermi momentum $p_F$, instead of trying to establish Cooper paring using BCS solution with EFT interactions\cite{Papen}. Again, we will pay more attention to the influences of $J_0$ and $x$ on the poles' existence and distribution.
\begin{figure}[ht]\begin{center}\resizebox{16cm}{!}{\includegraphics{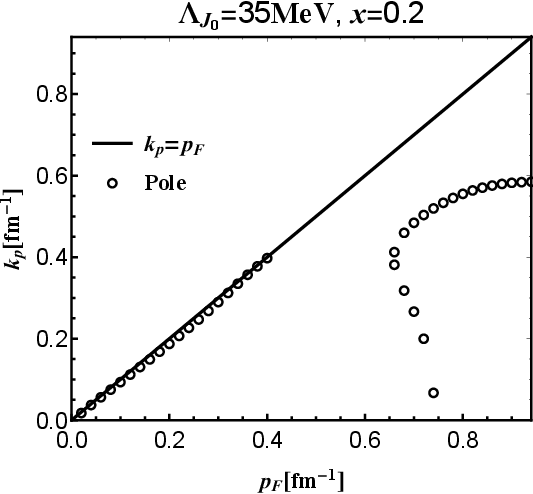}\includegraphics{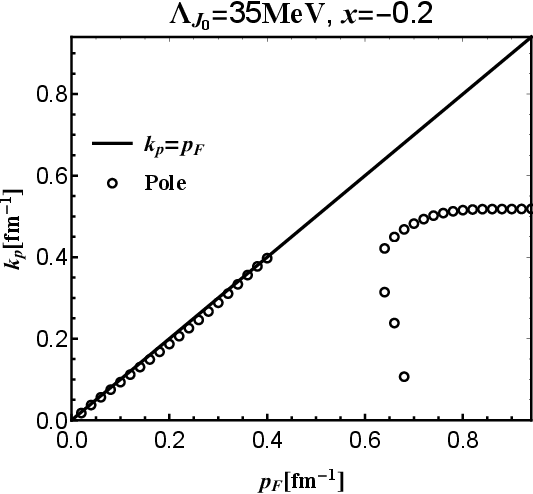}\quad\quad\includegraphics{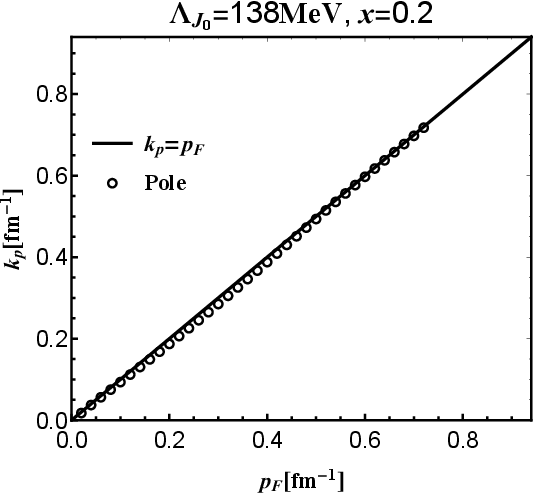}
\includegraphics{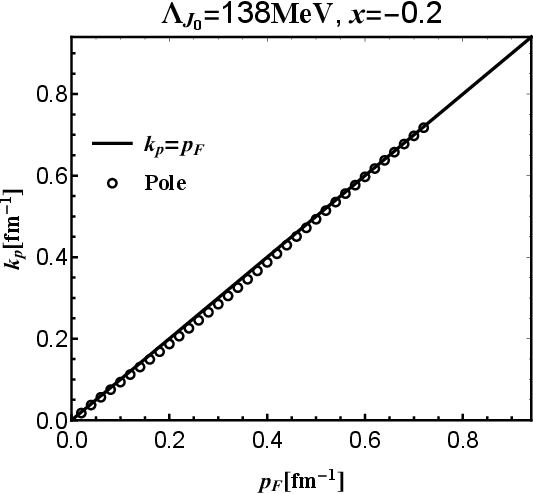}}\caption{Pole distributions of $G$ matrix below Fermi surface with difference choices of $x$ and $\Lambda_{J_0}$.}\end{center}\end{figure}
\begin{figure}[ht]\begin{center}\resizebox{16cm}{!}{\includegraphics{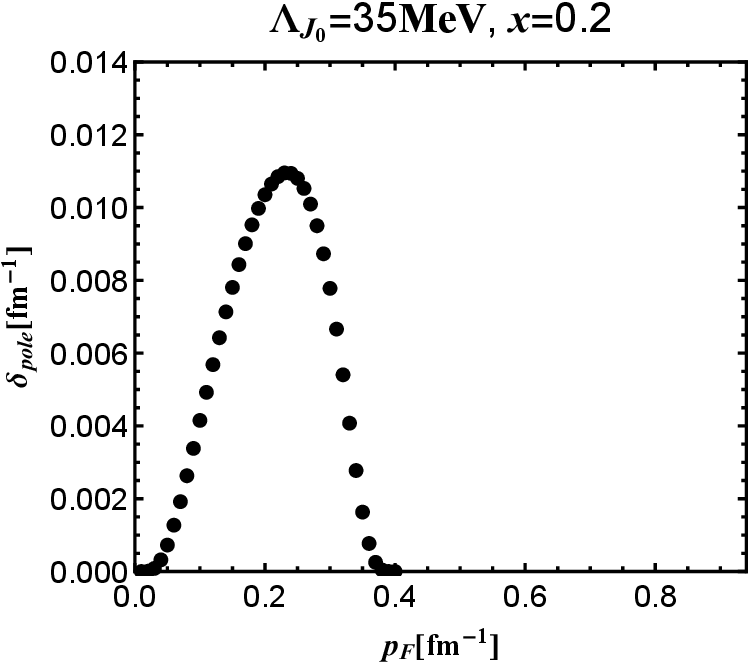}\includegraphics{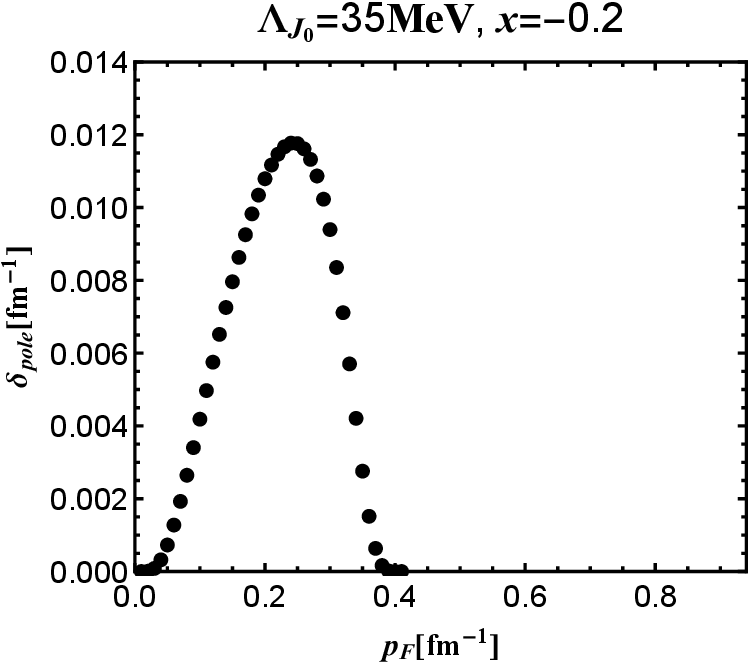}\quad\quad\includegraphics{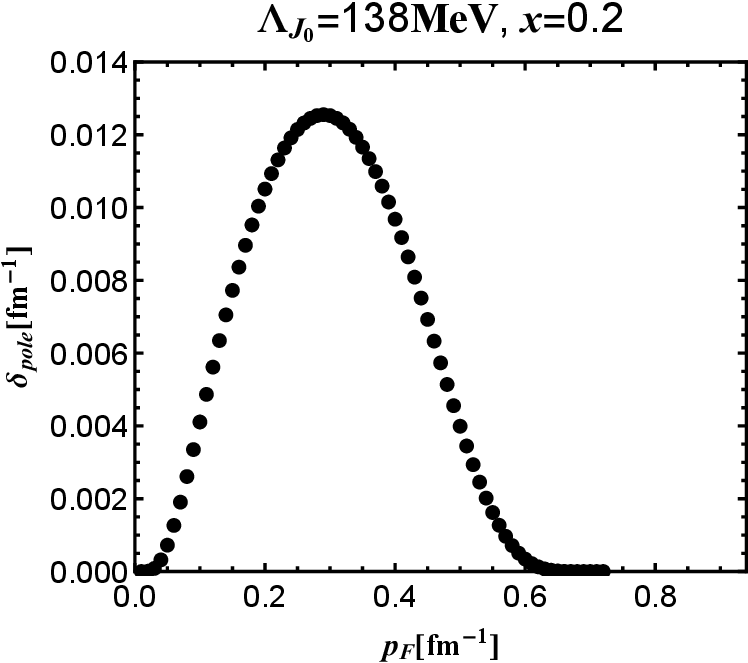}
\includegraphics{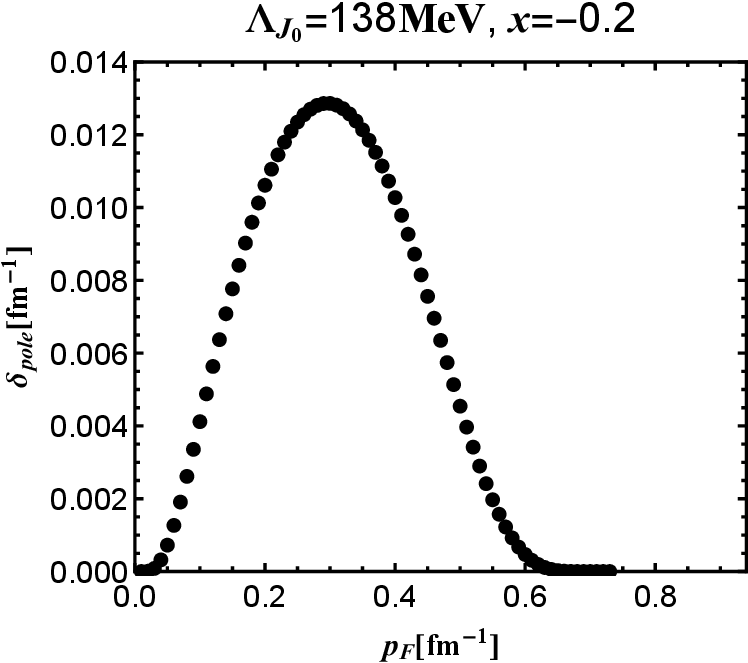}}\caption{The shape of 'differences' versus $p_F$ with difference choices of $x$ and $\Lambda_{J_0}$.}\end{center}\end{figure}

To proceed, we seek for the solutions to the equation $\frac{1}{G(k_p;p_F)}=0$ in terms of $k_p$ with a given $p_F$. At order $\mathcal{O}(Q^2)$, the explicit form of this equation reads\bea\label{pole}\frac{1}{\tilde{\alpha}_{s;0}(p_F)+\tilde{\alpha}_{s;1}(p_F)k^2_p}+J_0-\frac{M_Np_F}{2\pi^2}+\frac{M_Nk_p}{4\pi^2}\ln\frac{p_F+k_p}{p_F-k_p}=0,\eea with the density- or $p_F$-dependent $\tilde{\alpha}_{s;0,1}$ given in the \ref{appb}. Eq.(\ref{pole}) is solved numerically with choices $\Lambda_{J_0}=138$MeV, $35$MeV and $x=\pm0.2$. The real pole $k_p$ just below $p_F$, i.e., near Fermi surface, and the 'difference' $\delta_{pole}\equiv{p_F-k_p}$ are functions of $p_F$ and plotted in Fig.4 and Fig.5, respectively. It is obvious that $J_0$ and $\sigma_{s;2}$ do affect the detailed behaviors and even the existence of pairing in $^1S_0$ channel for a given $p_F$, and a smaller $\Lambda_{J_0}$ leads to a narrower window in $p_F$ allowing for a real near-Fermi-surface pole.

In particular, it can be shown that the pathological 'pole' behavior in energy curves exhibited in Fig.2 and Fig.3 stems from a pole term of second order in $G$ with $\Lambda_{J_0}=35$MeV: $G\propto(k-\alpha_{\pm})^{-2}$, where $\alpha_+\approx0.3996$fm$^{-1}$ for $x=0.2$ and $p_F=0.6588$fm$^{-1}$, $\alpha_-\approx0.3783$fm$^{-1}$ for $x=-0.2$ and $p_F=0.6289$fm$^{-1}$. In Fig.4 this is reflected by the fact that the pole solution disappears after $p_F\approx$0.4fm$^{-1}$ but reappears as a second-order pole that immediately bifurcates into two first-order poles around $p_F\approx0.6588(0.6289)$fm$^{-1}$. As these 'reappearing' poles lie 'far' below the Fermi surface, they could not be identified as the 'Cooper pairs'. In our view, such pathological 'form' of $G$ implies that $\Lambda_{J_0}=35$MeV is not a reasonable or physical choice for the pionless EFT description of nuclear matter, as in any many body system constituted by low energy nucleons the single particle energy value could not fluctuate from $\infty$ to $-\infty$ around a $p_F$ well below $\Lambda_{\not\pi}\approx0.699$fm$^{-1}$. By contrast, the $G$ with choice $\Lambda_{J_0}=138$MeV does not suffer from such pathological behavior as long as $p_F$ is less then 1.42fm$^{-1}$, a value about twice of the upper scale of pionless EFT. We should note that for $\Lambda_{J_0}=138$MeV the real near-Fermi-surface pole ceases to exist for about $p_F>0.68$fm$^{-1}$, a number quite close to the upper scale of EFT($\not\!\!\pi$): $\Lambda_{\not\pi}\approx0.699$fm$^{-1}$, again in contrast to the $\Lambda_{J_0}=35$MeV case where the corresponding value of $p_F$ is less than $0.4$fm$^{-1}$. The values of $\delta_{pole}$ in MeV are tabulated in Table 1 for $\Lambda_{J_0}=138$MeV, from which we see that they do take the typical values of the pairing gap in nuclear matter: around 2 MeV.
\begin{table}[!hbp]\label{gapvalue}\caption{$\delta_{pole}$ (MeV) versus $p_F$ (fm$^{-1}$) in $^1S_0$ at $\mathcal{O}(Q^2)$ with $\Lambda_{J_0}=138$MeV}
\begin{center}\begin{tabular}{ccccccc}\hline\hline\\ $p_F$&$0.1$&$0.2$&$0.3$&$0.4$&$0.5$&$0.6$\\\\\hline\\$\delta_{pole}(p_F;x=+0.2)$&0.81&2.07&2.47&1.91&0.79&0.07\\ \\$\delta_{pole}(p_F;x=-0.2)$&0.81&2.09&2.54&2.03&0.90&0.09\\\\\hline\hline\end{tabular}\end{center}\end{table}

Note that the shape of 'difference' $\delta_{pole}$ as a function of $p_F$ bears a strong resemblance to the gap function $\Delta(p_F)$ determined from $NN$ interactions\cite{gap1,gap2,gap3,gap4}, suggesting that there should be a correspondence between $\delta_{pole}$ and $\Delta(p_F)$. From our presentation, it is obvious that for each given value of $p_F$, there is a unique value $\delta_{pole}$ (provided that $J_0$ and $\sigma_{s;2}$ are physically determined). Then the correspondence between $\Delta$ (also uniquely defined for a give n$p_F$) and $\delta_{pole}$ should also be unique for a given $p_F$, thus, $\delta_{pole}$ could be seen as another measure of the gap function $\Delta=g(p_F)$. Formally, as long as $\Delta=g(p_F)$ and $\delta_{pole}=\tilde{g}(p_F;J_0,\sigma_{s;2})$ are single-valued functions, we have\bea\delta_{pole}=\tilde{g}\left(g^{-1}(\Delta);J_0,\sigma_{s;2}\right)=f(\Delta;J_0,\sigma_{s;2}).\eea Of course, for such correspondence to be valid, $J_0$ and $\sigma_{s;2}$ must be determined via certain plausible means, which is left as an open task. We hope what we demonstrated above could help to convince the community that $J_0$ and $\sigma_{s;2}$ (and the like) should be taken as physical parameters to be determined rather than as running ones, at least for closed-form objects like $G$ or $T$ matrices. We have at present no direct clue for a convenient determination of such physical parameters, further exploration is needed.
\section{Discussions and Summary}Now it is time for some remarks: From our presentation, it is obvious that the nonperturbative scenario of renormalization of pionless EFT for $NN$ scattering could be readily carried over to nuclear matter or many-body context, keeping the power counting depicted for few-body context intact. The Fermi momentum $p_F$ or the density background sets an external scale that could be counted either as a typical scale or an upper scale for the EFT description to make sense, which imposes no obstacle in nonperturbative context. The running couplings remain density-independent, hence conceptually more consistent with the fact that the EFT couplings are originally defined in vacuum context, but the medium environment does provide additional 'physical' inputs of information or outer constraints for the calibration of the running couplings and related quantities. Again, as in the zero-density case\cite{YH,AOP,1806}, the lower order EFT couplings' renormalization are affected by higher couplings that characterize shorter distance interactions, not the reverse, which may be conjectured to be a general scenario of renormalization of EFT in nonperturbative contexts.

Using the closed-form $G$ matrix obtained, the energy per particle is computed for both neutron matter and symmetric nuclear matter, demonstrating the physical relevance of $J_0$ and $\sigma_{s;2}$ in reasonable description of physical issues, like the location of saturation in terms of density. Their relevance is also illustrated in the paring phenomenon exhibited via the poles of the $G$ matrix as a function of on-shell momentum with a given $p_F$. It is also worth mentioning that both the analytical form of energy per particle obtained within certain plausible approximation and the numerical one without that approximation exhibit better behaviors than the pure perturbative ones that is valid only within a very small window due to the large scattering length.

In summary, we presented the closed-form $G$ matrix in $^1S_0$ channel obtained from BG equation using pionless EFT. Our experience here tells us that the nonperturbative solutions in our formulation are still tractable and trustworthy in many-body context, at least for nuclear matter. The $G$ matrix obtained is also applied to study some physical issues like energy per particle and pairing for both neutron and symmetric nuclear matter to demonstrate the physical relevance of certain parameters like $J_0$ and $\sigma_{s;2}$, along with the better behaviors derived from the nonperturbative formulation of $G$ matrix.
\section*{Acknowledgement}The authors are grateful to Hong-Xing Qi and Guo-Tu Shen for their very kind helps on numerical treatment. We are grateful to the anonymous referees for their reports that greatly improved the manuscript. This project is supported by the National Natural Science Foundation of China under Grant No. 11435005 and by the Ministry of Education of China.
\appendix
\section{Integrals}\label{appa}The integrals involved in this report are parametrized and defined as below ($E^+\equiv E+i\epsilon,\ p\equiv\sqrt{M_NE}$)\bea&&\tilde{\mathcal{I}}_{2n} \equiv\int\theta_F(k)\frac{d^3k}{(2\pi)^3}\frac{k^{2n}}{E^+-k^2/M_N}=-\tilde{\mathcal{I}}_0p^{2n}+\sum_{l=1}^n\tilde{J}_{2l+1}p^{2(n-l)},\\&&\tilde{\mathcal{I}}_0 =\tilde{J}_0+i\theta_F(p)\frac{M_N}{4\pi}p+\frac{M_N}{4\pi^2}p\ln\frac{p_F+p}{|p_F-p|}=\int^\infty_{p_F}\frac{k^2dk}{2\pi^2}\frac{1}{k^2/M_N-E^+},\\&&\tilde{J}_0\equiv M_N\int^\infty_{p_F}\frac{dk}{2\pi^2}=J_0-\kappa_1,\quad\tilde{J}_{2n+1}\equiv-M_N\int^\infty_{p_F}\frac{k^{2n}dk}{2\pi^2}={J}_{2n+1}+\kappa_{2n+1},\ (n\geq1),\\&&{J}_0\equiv M_N\int^\infty_0\frac{dk}{2\pi^2},\ {J}_{2n+1}\equiv-M_N\int^\infty_0\frac{k^{2n}dk}{2\pi^2},\ \kappa_{2n+1}\equiv{M}_N\int^{p_F}_0\frac{k^{2n}dk}{2\pi^2}=\frac{M_N}{2\pi^2}\frac{p^{2n+1}_F}{2n+1},\ (n\geq1),\\&&\theta_F(x)\equiv\theta(x-p_F),\eea where $J_0$ and $J_{2n+1}$ are prescription dependent before imposing any constraints.

In dimensional schemes, we should have $J_{2n+1}=0$ and $\tilde{J}_{2n+1}\neq0$ since $\tilde{J}_{2n+1}=J_{2n+1}+\kappa_{2n+1}$ that is, the definite piece $\kappa_{2n+1}$ should be kept intact. In the implementation of KSW scheme in $G$ matrix\cite{krippa}, this definite piece is totally removed. In a sense, such implementation of PDS in medium background is questionable.
\section{Renormalization group invariants of $^1S_0$ at order $\mathcal{O}(Q^2)$}\label{appb}First, let us recast the ratios $\tilde{\alpha}_{p;0,1}$ into the following forms\bea \label{al0}&&\tilde{\alpha}_{s;0}=\frac{\alpha_{s;0}+\sigma_{s;2}^2\kappa_5}{\left(1-\sigma_{s;2}\kappa_3\right)^2},\ \tilde{\alpha}_{s;1}=\frac{\alpha_{s;1}-\sigma_{s;2}^2 \kappa_3}{\left(1-\sigma_{s;2}\kappa_3\right)^2},\\&&\label{sig}\sigma_{s;2}\equiv\frac{C_{s;2}}{1-C_{s;2}J_3}.\eea Since $[\tilde{\alpha}_{p;0,1}]$ and $[\alpha_{s;0,1}]$ are physical parameters we have:\bea\mu{d}_{\mu}\left\{\tilde{\alpha}_{s;0},\tilde{\alpha}_{s;1}\right\}=0,\ \mu{d}_{\mu}\left\{\alpha_{s;0},\alpha_{s;1}\right\}=0,\eea where $\mu$ denotes a generic running scale possibly involved in any of the arguments of $[\tilde{\alpha}_{p;0},\tilde{\alpha}_{p;1}]$ as formal functions of couplings and $[J_{\cdots}]$. Then from Eq.(\ref{al0}) we have\bea\left(1-\sigma_{s;2}\kappa_3\right)^{-3}\left(\alpha_{s;0}+\sigma_{s;2}\kappa_3^{-1}\kappa_5\right)\mu{d}_{\mu}\sigma_{s;2}=0,\ \left(1-\sigma_{s;2}\kappa_3\right)^{-3}\left(\alpha_{s;1}-\sigma_{s;2}\right)\mu{d}_{\mu}\sigma_{s;2}=0.\eea The only possibility of the two equations will be that\bea\mu{d}_{\mu}\sigma_{s;2}=0.\eea It is a formal way of saying that Eqs.(\ref{al0}) forbid the 'running' of $\sigma_{s;2}$, a constraint due to the presence of density. This could also be derived from the renormalization of the off-shell closed-from $T$ matrix as mentioned in Sec.3.1.

Apart from $\sigma_{s;2}$, $J_3$ is also renormalization group invariant due to the identity\bea\label{identity}\alpha_{s;1}-2\sigma_{s;2}-\sigma_{s;2}^2J_3=0\eea that could be seen by combining the definitions in Eq.(\ref{alphazd}) and Eq.(\ref{sig}).
\section{Cancelation of $p_F$-dependence in the running couplings of $^1S_0$ channel}\label{appc}To proceed, we employ Eqs.(\ref{al0}), Eq.(A.3) and Eq.(\ref{identity}) to simplify the formal solutions in Eq.(\ref{RCtilde})\bea\tilde{C}_{s;2}&=&\frac{\sigma_{s;2}^2}{\alpha_{s;1}-2\sigma_{s;2}+\sigma_{s;2}^2\kappa_3} \left(1-\frac{1}{\sqrt{1+\frac{\alpha_{s;1}-\sigma_{s;2}^2\kappa_3}{(1-\sigma_{s;2}\kappa_3)^2}\cdot\frac{\alpha_{s;1}-2\sigma_{s;2}+\sigma_{s;2}^2\kappa_3}{\sigma_{s;2}^2}}}\right) \nonumber\\&=&\frac{\sigma_{s;2}^2}{\alpha_{s;1}-2\sigma_{s;2}+\sigma_{s;2}^2\kappa_3}\left(1-\frac{\sigma_{s;2}-\sigma_{s;2}^2\kappa_3}{\alpha_{s;1}-\sigma_{s;2}}\right)= \frac{\sigma_{s;2}^2}{\alpha_{s;1}-\sigma_{s;2}}=C_{s;2},\eea where we have chosen the $+$ root in order to reproduce the vacuum case in the limit $p_F\rightarrow0$. Similarly, we have\bea\tilde{C}_{s;0}&=&\frac{\alpha_{s;0}+\sigma_{s;2}^2\kappa_5}{(1-\sigma_{s;2}\kappa_3)^2}\left(\frac{\sigma_{s;2}-\sigma_{s;2}^2\kappa_3}{\alpha_{s;1}-\sigma_{s;2}} \right)^2-\frac{(J_5(\mu)+\kappa_5)\sigma_{s;2}^4}{(\alpha_{s;1}-2\sigma_{s;2}+\sigma_{s;2}^2\kappa)^2}\left(1-\frac{1}{\sqrt{1+\frac{\alpha_{s;1}-\sigma_{s;2}^2\kappa_3} {(1-\sigma_{s;2}\kappa_3)^2}\cdot\frac{\alpha_{s;1}-2\sigma_{s;2}+\sigma_{s;2}^2\kappa_3}{\sigma_{s;2}^2}}}\right)^2\nonumber\\&=&\frac{\alpha_{s;0}+\sigma_{s;2}^2\kappa_5} {(\alpha_{s;1}-\sigma_{s;2})^2}\sigma_{s;2}^2-\frac{J_5(\mu)+\kappa_5}{(\alpha_{s;1}-\sigma_{s;2})^2}\sigma_{s;2}^4=\frac{\alpha_{s;0}-\sigma_{s;2}^2J_5(\mu)} {(\alpha_{s;1}-\sigma_{s;2})^2}\sigma_{s;2}^2=C_{s;0}.\eea

The rationale for this property is that since the EFT couplings are originated from short-distance dynamics, then their renormalization could be only affected by the processes in vacuum and hence independent of medium environment.
\section{$G$ matrix and running couplings in uncoupled $P$ channels}\label{appd}The leading order ($\mathcal{O}(Q^2)$) uncoupled $P$ channel $G$ matrix is also quite simple:\bea&& \frac{1}{G_P(p;p_F)}=\frac{1}{\tilde{\alpha}_{P;0}p^2}+\tilde{\mathcal{I}}_0,\quad\tilde{\alpha}_{P;0}\equiv\frac{C_{P;2}}{1-C_{P;2}\tilde{J}_3}=\frac{\alpha_{P;0}} {1-\alpha_{P;0}\kappa_3},\ \alpha_{P;0}\equiv\frac{C_{P;2}}{1-C_{P:2}{J}_3}.\eea Simple as it is, here we have two renormalization group invariants, $J_0$ and ${\alpha}_{P;0}$ from which we have\bea{C}_{P;2}=\frac{\alpha_{P;0}}{1+\alpha_{P;0}J_3(\mu)}.\eea

At order $\mathcal{O}(Q^4)$, the $P$ channel $G$ matrix reads\bea&&\frac{1}{G(p;p_F)}=\frac{\tilde{N}_{P;0}+\tilde{N}_{P;1}p^2}{\tilde{D}_{P;0}p^2+\tilde{D}_{P;1}p^4} +\tilde{\mathcal{I}}_0,\\&&\tilde{N}_{P;0}=(1-C_{P;4}\tilde{J}_5)^2-C_{P;2}\tilde{J}_3-C_{P;4}^2\tilde{J}_3\tilde{J}_7,\ \tilde{N}_{P;1}=C_{P;4}^2\tilde{J}_3\tilde{J}_5 -2C_{P;4}\tilde{J}_3,\\&&\tilde{D}_{P;0}=C_{P;2}+C_{P;4}^2\tilde{J}_7,\ \tilde{D}_{P;1}=2C_{P;4}-C_{P;4}^2\tilde{J}_5.\eea Obviously, the following ratios must be renormalization group invariant,\bea\tilde{\alpha}_{P;0}\equiv\tilde{N}_{P;0}^{-1}\tilde{D}_{P;0},\ \tilde{\alpha}_{P;1}\equiv \tilde{N}_{P;0}^{-1}\tilde{D}_{P;1},\ \tilde{\beta}_{P;1}\equiv \tilde{N}_{P;0}^{-1}\tilde{N}_{P;1}.\eea Then we can obtain the running couplings simply by making the replacement $[J_{\cdots}\rightarrow\tilde{J}_{\cdots}]$ from that given in Ref.\cite{1806}\bea{C}_{P;4}=\tilde{J}_5^{-1}\left\{1-\sqrt{1-\frac{\tilde{\alpha}_{P;1}\tilde{J}_5}{1+\tilde{\alpha}_{P;0}\tilde{J}_3+\tilde{\alpha}_{P;1}\tilde{J}_5}}\right\},\ {C}_{P;2}=\frac{\tilde{\alpha}_{P;0}}{1+\tilde{\alpha}_{P;0}\tilde{J}_3+\tilde{\alpha}_{P;1}\tilde{J}_5}-\tilde{C}_{P;4}^2\tilde{J}_7.\eea Similar to the $^1S_0$ case, the apparent density-dependence cancels out after plugging in the full contents of renormalization group invariance, resulting in the following density-independent running couplings\bea{C}_{P;4}=\frac{\sigma^2_{P;4}}{\alpha_{P10}-\sigma_{P;4}},\ C_{P;2}=\frac{\alpha_{P00}-\sigma^2_{P;4}J_7(\mu)}{(\alpha_{P10}-\sigma_{P;4})^2}\sigma_{P;4}^2,\eea with \bea\alpha_{P00}\equiv\frac{\tilde{D}_{P;0}(p_F=0)}{(1-C_{P;4}J_5)^2},\ \alpha_{P10}\equiv\frac{\tilde{D}_{P;1}(p_F=0)}{(1-C_{P;4}J_5)^2},\ \sigma_{P;4}\equiv\frac{C_{P;4}} {1-C_{P;4}J_5}.\eea The renormalization group invariance of $\alpha_{P;00},\alpha_{P;10}$ and $\sigma_{P;4}$ will be shown in \ref{appe}.
\section{Renormalization group invariants in the uncoupled $P$ channels}\label{appe}First, let us recast $[\tilde{\alpha}_{P;i},\ i=0,1;\tilde{\beta}_{P;1}]$ into the following transparent form\bea\label{aptilde}\tilde{\alpha}_{P;0}=\frac{\alpha_{p00}+\sigma^2_{P;4}\kappa_7}{(1-\sigma_{P;4}\kappa_5)^2-\tilde{J}_3(\alpha_{P00}+\sigma^2_{P;4}\kappa_7)},\ \tilde{\alpha}_{P;1}=\frac{\alpha_{p10}-\sigma^2_{P;4}\kappa_5}{(1-\sigma_{P;4}\kappa_5)^2-\tilde{J}_3(\alpha_{P00}+\sigma^2_{P;4}\kappa_7)}.\eea

Second, $J_3$ is already renormalization group invariant due to the constraint $\tilde{\beta}_{P;1}+\tilde{J}_3\tilde{\alpha}_{P;1}=0$ which is ready to see from the expressions of $\tilde{N}_{P;1}$ and $\tilde{D}_{P;1}$ given above.

Now, consider $\mu{d}_{\mu}\{\tilde{\alpha}_{P;0},\tilde{\alpha}_{P;1}\}=0$, or equivalently $\mu{d}_{\mu}\left(\tilde{\alpha}^{-1}_{P;0}\right)=0,\ \mu{d}_{\mu} \left(\tilde{\alpha}_{P;1}\tilde{\alpha}^{-1}_{P;0}\right)=0.$ Bringing in the detailed expressions of $\tilde{\alpha}_{P;0,1}$ listed in Eq.(\ref{aptilde}), we can find from $\mu{d}_{\mu}\left(\tilde{\alpha}^{-1}_{P;0}\right)=0$ that\bea\left(\kappa_5-\sigma^{-1}_{P;4}\right)\mu{d}_{\mu}\left(\sigma_{P;4}^{-2}\alpha_{P00}\right)+2\sigma_{P;4}^{-2} \left(\sigma_{P;4}^{-2}\alpha_{P00}+\kappa_7\right)\mu{d}_{\mu}\sigma_{P;4}=0.\eea As terms of different powers of $p_F$ must be independent of each other, we conclude that\bea\mu{d}_{\mu}\sigma_{P;4}=0,\quad\mu{d}_{\mu}\alpha_{P00}=0.\eea Then, from $\mu{d}_{\mu}\left(\tilde{\alpha}_{P;1}\tilde{\alpha}^{-1}_{P;0}\right)=0$, we have\bea\mu{d}_{\mu}\alpha_{P10}=0.\eea Just like $J_3$ in Eq.(\ref{identity}), here $J_5$ becomes 'physical' due to similar reason:\bea\alpha_{P10}-2\sigma_{P;4}-\sigma^{2}_{P;4}{J}_5=0.\eea

\end{document}